# Cosmology holography the brain and the quantum vacuum

Antonio Alfonso-Faus

Departamento de Aerotécnia

Madrid Technical University (UPM), Spain

February, 2011. E-mail: aalfonsofaus@yahoo.es

**Abstract:** Cosmology, as a science, started at the beginning of the last century with the advent of the Einstein cosmological equations. Based on these equations, the present main stream cosmological model is the well known big-bang, this name unwillingly coined by Fred Hoyle many years ago. Relatively recent additions to this model have been inflation, dark matter and dark energy. We present a smoothly behaved new cosmological model that mainly takes into account the dark part of the Universe. We consider it as the background frame, the substrate, of what we see. The inclusion of the holographic principle clarifies the entropy problem that we also apply to the human brain. We take it as an engineering information center. Finally, the inclusion of the quantum vacuum in this scene creates an important challenge, an opportunity for future research in the knowledge of the Universe.

**Keywords:** Cosmology, information, entropy, gravitation, holographic principle, fractality, human brain, vacuum energy.

## 1. – Introduction

The subject of cosmology is a fascinating one: almost any one knows the basic ideas of the big-bang model. Present accelerated expansion of the universe is very well known too. Perhaps inflation is not so much, but here we present a new method to short out important discrepancies in this field: inflation (yes or no?), expansion (yes or no?), lifetime of the universe (finite or infinite?). The method is to use the measured values of the so called deceleration parameter $q$ and arrive at the history of the size of the universe.

The entropy issue contains intriguing questions on information, the unit bit and its physical interpretation. The equality of very large dimensionless numbers shows light on this issue here. We also arrive at the conclusion, by means of their definitions, that the universe follows the holographic principle and that is fractal.

Finally we address two important issues: the brain information capacity and the vacuum information content. The importance of the self gravitational potentials is here proposed as a very interesting field of research for the future.

## 2. - The cosmological odyssey

### 2.1 The static universe

About 100 years ago cosmology, as a science, started with the work of Einstein, with his field equations of general relativity applied to the universe. From then on an odyssey developed and is still going on today. To begin with, the attractive force of gravity alone would most probably have collapsed the universe in a rather short time. Einstein realized this and introduced his cosmological constant $\Lambda$, as a pushing force to balance gravitation. In this way the static model of the universe was brought to equilibrium.

### 2.2 The expanding universe

The Hubble´s interpretation of his observations of the red shifts from distant galaxies gave birth to a new idea: that the universe was expanding. Going back in time, as a thought experiment, the universe would be seen contracting. The known physics of gases immediately implied that the temperature and density of the universe would become higher and higher going to the past. The mathematical extrapolation to the first stages of the universe clearly gave the idea of an initial ~ infinite value for its temperature and density, a singularity. This initial "explosion" was unwillingly coined as a big-bang by Fred Hoyle.

### 2.3 The big-bang versus the steady state model

The expansion of the universe was generally accepted as a sound conclusion from the Hubble´s observations. Nevertheless the 3 scientists Hoyle, Bondi and Gold developed the idea of a continuous creation of matter, together with the accepted expansion, in such a way that all physical properties remained constant in time, creation compensating for expansion. Then a steady state model was brought into the picture as an alternative to the big-bang. In time the steady state model was generally abandoned, mainly because it had no evolution that was very well established. However, the big-bang model had many drawbacks. One of them, a very serious one, was that it could not explain the present size of the universe. Then a new idea came into the picture to solve this problem: "inflation".

### 2.4 Inflation

The two scientists, Allan Guth [1] and Linde [2], introduced the idea of an exponential expansion during a very short time, an initial inflation related to the vacuum properties (may be of space-time), and helped to increase the size when followed by the "normal" expansion. Today this hypothesis is getting more and more support due to the agreements seen with its predictions: flat universe, critical density, properties of the cosmic background radiation and so on. Usually this inflation epoch is maintained to have occurred after the "big-bang". Personally I do not see any need to day for the survival of the big-bang idea: inflation from an initial quantum black hole is enough to do the job. I refer to the fact that the Planck´s quantum black hole physical properties (mass, length, time and so on), when multiplied by the dimensionless scale factor $10^{61}$, give the present physical properties of our universe!

2.5 Accelerated expansion

The picture of the history of the universe, as we are presenting here, is that of a sudden and fast exponential expansion (initial inflation) followed by a much slow "normal" expansion. The point is that in order to break the fast exponential expansion gravitational attraction has to enter into the picture. Something like artificial fireworks that explode in an accelerated expansion followed by more expansion but decelerated. This picture would give as of today a decelerated expansion for the universe. If *a(t)* is the cosmological scale factor, its first derivative *ȧ(t)* will give the speed of expansion, and its second derivative *ä(t)* the acceleration, if it is a positive function of time (*ä(t)*>0), or the deceleration if it is a negative function of time (*ä(t)*<0). The "deceleration" parameter *q* was defined as

$$q(t) = - ä(t)\, a(t)/ ȧ(t)^2 \qquad (1)$$

One would expect that, after inflation, the universe was left still expanding but slowly decelerating. Then the definition of *q* in (1) would give (*ä(t)* < 0) a value of *q > 0*. This was the reason for including a negative sign in (1), to get a positive value for *q*, even to call this *q* the deceleration parameter, a positive dimensionless number. But nature keeps giving us surprises. About ten years ago it was found that the universe is in a state of *accelerated expansion,* and we know today that this state of acceleration started back in time at about one half the age of the universe. We needed the cosmological constant Λ to balance gravitation, to have a push to explain the expansion of the universe. This pressure obviously implies energy, and the constant Λ has been linked with this energy. In many instances it has been identified with the vacuum energy. But a big problem is still present, a very big one indeed. There is a huge factor of difference (about $10^{120}$) between the cosmological approach for Λ and the vacuum energy one. In the last section we give a new approach with the introduction of the self gravitational potential energies.

To have a measure of the expansion rate of the universe, a well known parameter H (the Hubble parameter) was defined as

$$H(t) = \dot{a}(t)\, a(t) \qquad (2)$$

The combination of the definitions (1) and (2) give the important cosmological equation

$$\dot{H} + [1 + q(t)]H^2 = 0 \qquad (3)$$

The exponential expansion introduced by Guth and Linde, inflation, can be interpreted as a very short period of time such that the Hubble parameter H is constant with time. And this must have occurred very close to the initial stages of the universe. From (3) we see that this condition implies for *q* the constant initial value *q = -1*. Given that this must have occurred during a very short time, as is the case for inflation, the expected value for *q* after inflation, and very near the initial stages, should be close to zero. The present value of *q* is about $q_0 \approx -0.67$. In between this two stages ( t ≈ 0 and today $t_0 \approx 1.37\ 10^{10}$ years) one must have a value of zero for *q*. This is because one must cross from a positive value (deceleration after inflation) to a negative one, an acceleration starting at about t ≈ 0.42 $t_0$ ( *with q = 0* ) and going on up to the present time ($q_0 \approx -0.67$). With these three values of *q:* 0, 0, and -0.67 corresponding to the ages 0, 0.42 $t_0$, and $t_0$ we can make an ASANTZ approach adjusting a parabolic curve with vertical axis:

$$q(t) = -1.1552\ x^2 + 0.4852\ x \qquad (4)$$

where the dimensionless parameter $x$ is defined as $t/t_0$. With this function of $q(t)$ so adjusted we can integrate the differential equations (3) and (2) to obtain the cosmological scale factor $a(t)$. We have used the present value of $H_0 t_0 \approx 1$ in order to define the limits of integration, as well as the definition $a_0 = a(t_0)$. In the Fig. 1 below we have the plot of $a(t)/a_0$ in the vertical axis versus the parameter $x = t/t_0$ in the horizontal axis.

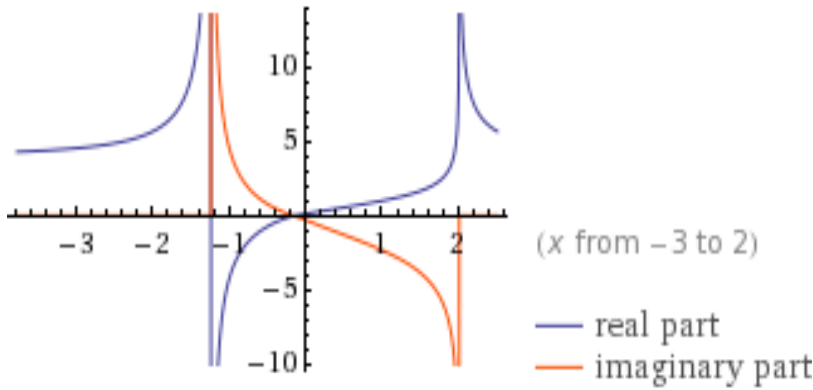

Fig. 1 Cosmological scale factor $a(t)/a_0$ versus time $x = t/t_0$

This plot of $a(t)$ has a lot of information in it. First we see an almost linear expansion from $x = 0$ to a little more than $x = 1$ (today). Second we see that there is an imaginary part of the plot that is negative (between $x = 0$ and $x = 2$). As we will see this strongly suggest the splitting between positive rest mass and negative imaginary mass, as we will see later. Third, there is no such splitting after $x = 2$. And last, but not least, we have two vertical asymptotic behaviors, an infinite for $a(t)$ at about $x = 2$, twice our present age, and another event before the first inflation ($x \approx 0$) at $x \approx -1.2$   In the next sections, 2.6 and 2.7, we give an interpretation to these results.

### 2.6 The end of the universe

Looking at Fig. 1 for $x = 2$ we see that the universe has spread to infinity at that age. This is an expected extrapolation from the accelerated expansion seen today that goes on until $a(t) \to \infty$ . It gives a total lifetime for our universe $t_f \approx 2\,t_0 \approx 2.74\ 10^{10}$ years.

Today we are at about 50 % of the lifetime. In Fig.2 we have the speed of expansion versus time

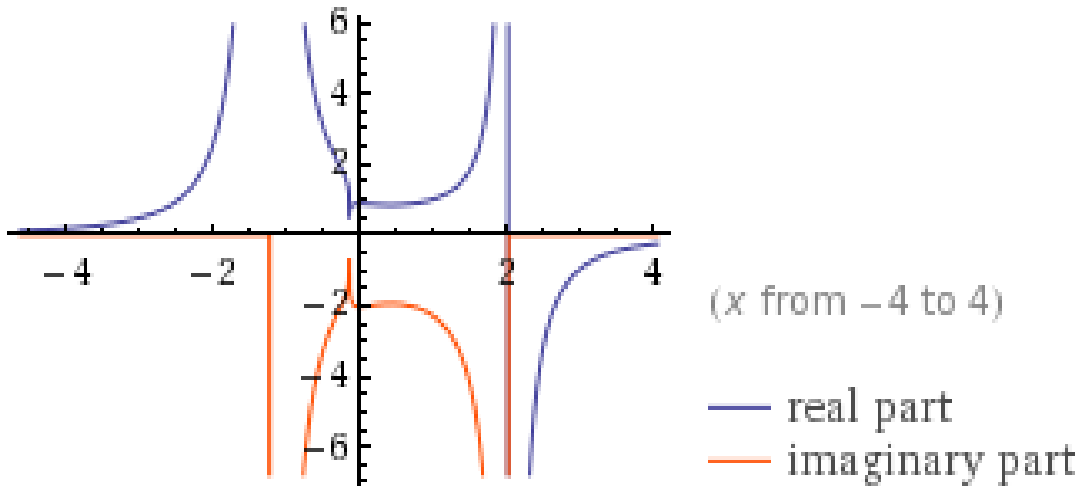

Fig. 2 Vertical axis: speed of expansion relative to the speed of light c. Horizontal axis: time t relative to the present age, $x = t/t_0$.

We see in Fig.2 that the speed of expansion, relative to the speed of light, is about 1 from the beginning to a little more than $x = 1$ (today). From then on it grows up to an infinite value at the end of the life of our universe, for $x = 2$.

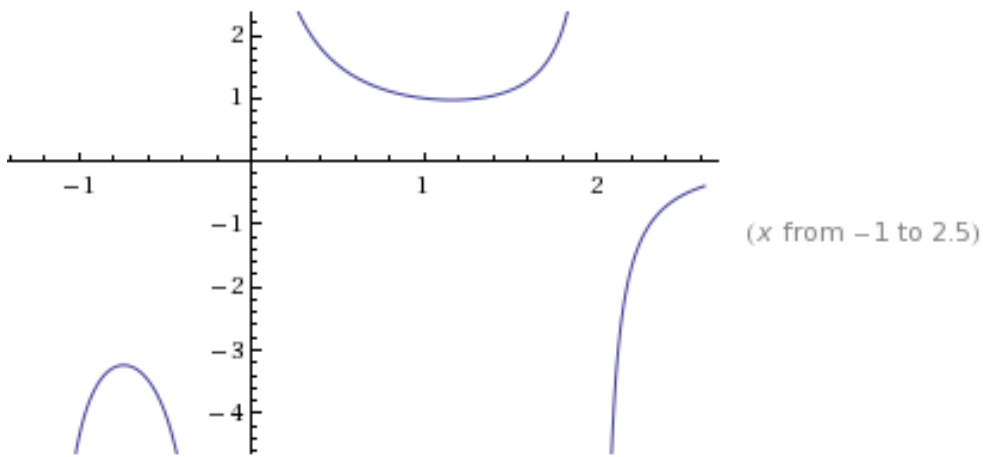

Fig. 3 Vertical axis: dimensionless Hubble parameter $Ht_0$ versus age, $x = t/t_0$, at the horizontal axis. There are two infinites for H: one at $x \approx 0$, the initial inflation, and the last one at the end of the universe, $x = 2$.

In Fig 3 we see the evolution of the Hubble parameter h with time. The initial inflation is evident, clearly giving the exponential expansion. The final "inflation" is also evident, showing the disaggregation of everything in our universe. This is an inevitable conclusion. At x = 2 the infinite value for the cosmological scale value in Fig.1, the infinite value for the expansion of the universe in Fig. 2 and the infinite value for the Hubble parameter H in Fig. 3 show the total disaggregation in our universe. If we follow Mach´s ideas on the inertial and gravitational masses, the interaction of any mass m with the rest of the universe is clearly zero. Then there is no mass m and therefore the disaggregation is total and at all levels. It is the end of our universe as we know it.

2.7 Black hole formation before x = 0.

Looking at the Figures 1, 2 and 3 it is clear that at an age x < 0 there is structure in the graphs suggesting the possibility of black hole formation by gravitational collapse. There are some theoretical comments in this direction in the literature. Here we just point out that research in this field may have sense.

## 3. – Information, the laws of physics and matter

3.1 Entropy of the universe

We can think of our universe as a kind of "quantum" black hole [3] and apply the Hawking-Bekenstein [4] and [5] formulation for its entropy S:

$$S = 4\pi \, k/\hbar c \, GM^2 \qquad (5)$$

Using the black hole relation between its mass M and its size *a(t)*

$$2GM/c^2 = a(t) \qquad (6)$$

And combining (5) and (6) we get (with the linear relation *a(t) ≈ ct)*

$$S = 4\pi \, k/\hbar c \, G(c^3 t/2G)^2 = \pi \, k/\hbar \, (c^5/G) \, t^2 \qquad (7)$$

And in natural units $G = c = \hbar = k = 1$ we finally get

$$S = \pi \, t^2 \qquad (8)$$

Going on using natural units, in Planck´s units of time we have then from (8)

$$S \approx 10^{122} \qquad (9)$$

The entropy of the universe increases with time and will arrive at a maximum at $x = 2$, its lifetime, and has a value of the order of $\sim 10^{122}$.

### 3.2 Entropy and gravity: the mass-energy of information

The quantum of gravity with mass $m_g$ has been presented [6] as

$$m_g = \hbar/(c^2 t) \approx 10^{-65} \text{ grams} \qquad (10)$$

Since the mass of the universe M is about $10^{56}$ grams, one has the number of gravity quanta $N_g$ in the universe as

$$N_g \approx M/m_g \approx 10^{122} \qquad (11)$$

The two very large numbers in (9) and (11), being of the same order of magnitude, give us a very strong reason to believe that the entropy S of the universe is the number of gravity quanta, as proposed 10 years ago [6], and this is the number of bits I it contains:

$$I \approx S \approx N_g \approx M/m_g \approx 10^{122} \qquad (12)$$

We now make a further and very intriguing conjecture:

*The unit of information, the bit, has a mass $m_g$ and an energy $m_g c^2 \approx \hbar/t$, i.e. it is the same as the quantum of gravitational energy $\hbar\omega \approx 10^{-45}$ ergs. The total number of bits in the universe is then of the order of $10^{122}$.*

## 4. – The holographic principle and fractality

### 4.1 The holographic principle and information

The holographic principle, [7] and [8], basically states that the information contained in a closed volume V, with surface boundary A, is equivalent to the information contained in the bounding surface A, when using the Planck´s unit of surface to express A. Applied to the universe it implies that the information it contains is equivalent to the area of its event horizon, considering the universe as a black hole. In Planck´s units, the so called natural system, the area of the event horizon for the universe is $4\pi t^2 \approx 10^{122}$ (see (8) and (9)), which is the Hawking-Bekenstein entropy formulation. Then, the amount of information in the universe is of the order of its entropy, about the number of gravity quanta, the number of bits, and the area of the event horizon in Planck´s units. We see that the mass of the universe, as given by the amount of information it contains, is about $10^{122}$ $m_g$ $\approx 10^{56}$ grams.

### 4.2 Fractality

Any mass M inside a volume of size R is said to have the fractal property of dimension D if for any value $M_i$, and size $R_i$, one has the relation

$$M_i/R_i^D = \text{constant} \tag{13}$$

The universe has the fractal property, or very close to it, for D = 2. We can check it for the whole universe with M $\approx 10^{56}$ grams, and R $\approx 10^{28}$ cms. With these values the constant in (13) is just 1gram/cm$^2$. This is very close to the value for any system in the universe. For example galaxies, protons, etc. The fractality of the universe can then be expressed as

$$M_i/R_i^2 = \text{constant} \qquad (14)$$

Applying the holographic principle we have that the total information contained in the volume of size R is equal to its mass M in units of gravity quanta, i.e., $10^{122}$ bits. And this is the same as its event horizon area expressed in units of Planck´s. In fact we see here a kind of equivalence between the holographic principle and fractality, for D = 2. Then we make the following conjecture

*The holographic principle applied to a 3 dimensional volume of mass M, and using the unit mass of the bit $m_g$, is equivalent to the fractal property condition, expressing the area in Planck´s units. The universe is holographic and fractal.*

## 5. – The human brain

### 5.1 Information capacity

Given the mass of a system m the number of bits it contains is just $m/m_g$. Then the number of bits that the human brain may contain is about $10^{69}$. This is equivalent, using the holographic principle, to the area of its *cortex*, the boundary of the human brain, in Planck´s units. The similarity between the human brain and the whole universe is ensured by the holographic principle. In the next section we will see that the bit information is related to the quantum vacuum.

## 6. – The Λ problem and the vacuum energy

### 6.1 The Λ problem

When dealing with cosmology, using the Einstein´s cosmological equations, one gets an order of magnitude for the Λ "constant" ~ $1/t^2$. When dealing with the standard model for fundamental particles one gets a factor of about $10^{122}$ for the corresponding constant. This enormous difference has been explained [9] as a scale factoring, universe versus Planck´s scale. Here we can relate now this $\Lambda$ problem to the vacuum energy, the quantum vacuum. We do this in the next section.

### 6.2 The vacuum energy

Mach´s principle can be stated as saying that the rest energy of a mass m is equal to the gravitational potential energy of this mass with respect to the rest mass M of the universe:

$$mc^2 = GMm/R \tag{15}$$

Within a factor of two this is equivalent to say that the gravitational radius of the universe, $GM/c^2$, is just its size R, a kind of black hole. Another way to look at this is to say that the self gravitational potential energy of the universe is equal to its relativistic rest energy

$$GM^2/R = Mc^2 \tag{16}$$

Multiplying by the factor $10^{-122}$ we transform (16) to

$$GM\,m_g/R = m_g c^2 \tag{17}$$

Then using the product equality $M\,m_g \approx 10^{39}\,m_{fp}^2$, where $m_{fp}$ is the mass of a fundamental particle, and the scale relation $R \approx 10^{39} r_{fp}$, where $r_{fp}$ is its size, we transform (17) to

$$Gm_{fp}^2/r_{fp} \approx m_g c^2 \tag{18}$$

This means that the self gravitational potential energy of a fundamental particle of mass $m_{fp} \sim 10^{-24}$ grams and size $r_{fp} \sim 10^{-12}$ cm is of the same order of magnitude as the energy of one bit. It is now clear that the vacuum energy conclusions from the standard model, predicting a $\Lambda$ value off by a huge factor, $10^{122}$, is due to the use of rest energy of masses instead of their self gravitational potential.

## 7. – Conclusions

We have proved here that one needs only one cosmological parameter to find the cosmological scale factor, the size of the universe in terms of time. It is the deceleration parameter $q$, (and the present value of the Hubble parameter $H_0$). We obtain the complete history of the universe, initial inflation, deceleration, acceleration and final expansion to infinity at twice the present age of the universe. This has been achieved without the use of the Einstein cosmological equations (based on his field equation of general relativity).

We have also seen the entropy dependence with cosmological time, and the identification of the unit of information, the bit, with the quantum of the gravitational field. Besides, the universe seems to be holographic and fractal.

The information capacity of the brain is found to be about $10^{69}$ bits. A similarity is suggested in the way the universe, as a black hole, can be described with a horizon and the information capacity in this surface: the cortex of the brain seems to have most of its information capacity.

Finally, the vacuum energy concept many times related to the cosmological constant $\Lambda$, may be substantiated if using the self gravitational potential energies instead of the total relativistic $mc^2$.

## 8. – Acknowledgements

## 9. – References